# Metamagnetic multiband Hall effect in Ising antiferromagnet ErGa$_2$

Takashi Kurumaji[1,2,*], Shiang Fang[1,3], Linda Ye[2], Shunsuke Kitou[4], and Joseph G. Checkelsky[1*]

[1] Department of Physics, Massachusetts Institute of Technology, Cambridge, MA 02139, USA.

[2] Division of Physics, Mathematics and Astronomy, California Institute of Technology, Pasadena, California 91125, USA.

[3] Department of Physics and Astronomy, Center for Materials Theory, Rutgers University, Piscataway, New Jersey 08854, USA.

[4] Department of Advanced Materials Science, University of Tokyo, Kashiwa 277-8561, Japan.

*Corresponding author: kurumaji@caltech.edu, checkelsky@mit.edu

**Abstract**

Frustrated rare-earth-based intermetallics provide a promising platform for emergent magnetotransport properties through exchange coupling between conduction electrons and localized rare-earth magnetic moments. Metamagnetism, the abrupt change of magnetization under an external magnetic field, is a signature of first-order magnetic phase transitions; recently, metamagnetic transitions in frustrated rare earths intermetallics have attracted interest for their accompanying nontrivial spin structures (e.g. skyrmions) and associated non-linear and topological Hall effects. Here, we present metamagnetism-induced Hall anomalies in single-crystalline $ErGa_2$, which recalls features arising from the topological Hall effect but wherein the strong Ising type anisotropy of Er moments prohibit noncoplanar spin structures. We show that the observed anomalies are neither due to anomalous Hall effect nor topological Hall effect, instead, can be accounted for via 4f-5d interactions which produce a band-dependent mobility modulation. This leads to a pronounced multiband Hall response across the magnetization process— a metamagnetic multiband Hall effect that resembles a topological-Hall-like response but without nontrivial origins. The present findings may be of general relevance in itinerant metamagnetic systems regardless of coplanar/non-coplanar nature of spins and are important for the accurate identification of Hall signals due to emergent magnetic fields.

**Significance Statement**

The topological Hall effect (THE), often assigned to jumps in Hall resistivity at metamagnetic transitions, is now widely employed as a tool to electrically probe emergent magnetic fields associated with spin textures such as skyrmions in conducting systems. An empirical ansatz is often used to experimentally extract the THE from the Hall responses; here, we present a Hall effect study of a frustrated triangular-lattice magnet $ErGa_2$ as a counter-example that suggests the need for more thorough consideration of this approach. The proposed multiband mechanism for the observed Hall anomaly is anticipated to be relevant to a wide class of magnetic systems. Our study sheds light on an overlooked transport effect and highlights crucial considerations when identifying nontrivial Hall responses.

**Main Text**

**Introduction**

The Hall resistivity $\rho_{yx}$ of noncoplanar magnetic metals is often empirically expressed as [1,2]

$$\rho_{yx} = \rho_{yx}^{O} + \rho_{yx}^{A} + \rho_{yx}^{T} = R_0 B + R_s M + \rho_{yx}^{T}, \qquad (1)$$

where the first term is the ordinary Hall effect (OHE) due to Lorentz force and the second term is the anomalous Hall effect (AHE) proportional to the magnetization ($M$). The third term is the topological Hall effect (THE) that appears *e.g.* when noncoplanar spin alignments in an intermediate field region are realized. Such spin textures give rise to finite scalar spin chirality, $\chi_{ijk} \propto \mathbf{S}_i \cdot (\mathbf{S}_j \times \mathbf{S}_k)$ [3-6], where $\mathbf{S}_{i,j,k}$ are spin moments at neighboring sites, $i$, $j$, $k$. $\chi_{ijk}$ acts as a local emergent magnetic field for conduction electrons to give rise to a transverse response [7-12]. To isolate these terms, in the simplest case $R_0$ and $R_s$, are characterized by measurements beyond magnetization saturation [13-15], with subsequent analysis of $\rho_{yx}^T$ involving the subtracting background $R_0 B + R_s M$. However, difficulty in isolating the topological Hall effect has recently been discussed in thin film systems [16-17]; the degree to which this is relevant for bulk crystalline systems has heretofore not been experimentally addressed.

Here, we report the Hall response in bulk single crystals, where the ansatz Eq. (1) breaks down. Our target compound is the frustrated magnet $ErGa_2$ [18] that exhibits a remarkable non-monotonic Hall response across metamagnetic transitions, but finite scalar spin chirality is forbidden due to strong Ising-type anisotropy. We propose a model that describes this as a multiband modulation of $R_0$ in the magnetization process, *i.e.* a metamagnetic multiband (MM) Hall effect. This generates a significant nonmonotonic field dependence of $\rho_{yx}$ arising from band-dependent spin-electron scattering, even without considering nontrivial AHE/THE terms. This model can generally be applied to multiband, metallic magnets. We anticipate the MM Hall effect can in principle occur in wider classes of systems potentially possessing noncoplanar spin structure alongside the AHE and THE therein. The finding of the present study highlights an important consideration when applying Eq. (1) to identify emergent magnetic field in frustrated itinerant magnets.

**Results**

We first summarize the model for the MM Hall response. In the presence of more than one type of conduction electrons, the coefficient $R_0$ in Eq. (1) in principle should reflect a multiband response; here we consider a representative two-band scenario with hole and electron pockets (their respective carrier densities as $n_h$ and $n_e$ and mobilities as $\mu_h$ and $\mu_e$). In the low-field regime ($\mu_{e,h}B < 1$), this is approximately [19]

$$R_0 = \frac{\rho_{yx}^0}{B} = \frac{1}{e}\frac{n_h\mu_h^2 - n_e\mu_e^2}{(n_h\mu_h + n_e\mu_e)^2}. \quad (2)$$

However, for magnetic systems the mobility is further modified by band-dependent spin-electron coupling. When the magnetization rapidly evolves at a metamagnetic transition this can give rise to the MM Hall effect. Importantly, in typical magnetically textured system (in particular those based on rare earth), metamagnetic transitions can take place at a relatively low magnetic field scale and can often be lower than the threshold for typical orbitally driven multiband (OHE) nonmonotonic Hall responses.

For concreteness, we consider a metallic compound with an electron band with 5d/6s orbital characters from magnetic cations and a hole band mainly composed of p-orbital at nonmagnetic (anionic) sites (see Fig. 1A). For the case of rare earth intermetallics, the local magnetic moment has the form of 4f levels that strongly couples with the cation conduction band through the intra-atomic 5d-4f interaction, $J\mathbf{S}\cdot\mathbf{s}\delta(\mathbf{r} - \mathbf{R})$ [20], where $J$ is exchange coupling constant, $\mathbf{S}$ and $\mathbf{s}$ are spins localized at $\mathbf{R}$ and for conduction electrons with spatial coordinate $\mathbf{r}$, respectively. The intersite interaction with the anion p-valence band, however, is expected to be relatively weak due to their reduced hybridization [21-23] and energy cost for 4f-p electron hopping (we consider the case where the 4f level is well-localized below the Fermi energy, which is typically realized in rare earth intermetallics apart from Ce, Pr, and Yb; see SI Sec. S1 for the opposite extreme where interatomic f-p hybridization becomes relevant). In such cases, the electron type carriers are more readily scattered from magnetic disorder (see Fig. 1B), giving rise to a field modulation in Eq. (2) via relative changes in $\mu_h$ and $\mu_e$.

This can be quantitatively illustrated with a simplified two-band model, in which we assume that $\mu_e$ is sensitive to while $\mu_h$ is immune from magnetic disorder scattering. Within Matthiessen's rule, the mobility of electron-type carriers is

$$\mu_e(B) = \frac{e}{m_e}\left(\frac{1}{\tau_0^e} + \frac{1}{\tau_{mag}^e(B)}\right)^{-1}. \quad (3)$$

The inverse of the carrier lifetime is additive: $\tau^{-1} = \tau_0^{-1} + \tau_{mag}^{-1}$, where $\tau_0$ is for the nonmagnetic (field-independent) scattering due to impurities/defects and phonons, and $\tau_{mag}$ stems from electron-spin scattering. In the magnetization process from the paramagnetic state (PM) to the field-induced ferromagnetic state (Fi-FM), the field-dependence of $1/\tau^e_{mag}$ can be expressed as

$$1/\tau_{mag}^e(B) = 1/\tau_{mag0}\,(1 - M(B)^2/M_0^2), \quad (4)$$

where the spin-disorder scattering is taken to be proportional to the magnitude of disorder upon uniformly aligned spins, *viz* $(1 - M^2/M_0^2)$ [24-26]; this provides a good description for Ising magnets wherein spin-flip and spin-wave scattering processes are minimal [27-29].

Figures 1C-E show the evolution of $\mu_e$, two-band resistivity ($\rho_{xx}$), and Hall resistivity ($\rho_{yx}$) for a linear evolution of $M$ as a general illustration for the case of constant $\mu_h$ and $n_{e,h}$. The negative magnetoresistance at low fields is consistent with field-suppression of magnetic disorder (spin MR) while the Hall resistivity exhibits a non-linearity. We note that this model does not include a contribution from $M$-linear AHE and THE and carrier density change. The modulation of the Hall effect arises from the field-dependence of OHE across the magnetization process. This effect

appears to be often overlooked in the conventional treatment of the anomalous Hall effect caused by side jump, skew scattering, and Berry phase mechanism, where the Hall anomaly originates from the second term in Eq. (1) and $R_0$ is taken as a constant.

The extension of this model to a metamagnetic transitions among antiferromagnetic (AFM), metamagnetic (M), and Fi-FM states is shown in Figs. 1F-H. At zero temperature, spin-disorder scattering or spin-wave scattering do not contribute to the spin MR. Instead, the order parameter and magnetic unit cell change across the metamagnetic transition result in a reconstruction of the magnetic Brillouin zone (BZ) and give rise to a stepwise change of the renormalized effective mass of carriers. The magnetoresistivity and Hall resistivity (Figs. 1G-H) both reflect the characteristic field dependence of $\mu_e$—the MM Hall effect. We note such affects ascribed to changes in carrier lifetime have been previously reported at metamagnetic transitions for other Ising-like rare earth systems [30-31]. The crucial point of this effect is that it should be distinguished from the topological Hall effect to avoid the overestimation of the emergent magnetic field (and in turn interpretation in terms of a noncoplanar nature of the underlying magnetic structure).

Here, we present a pronounced MM Hall effect in ErGa$_2$ as a remarkable example of the above scenario. Rare-earth intermetallics of the type $R$Ga$_2$ ($R$: rare earth) are frustrated magnets with an alternative $c$ axis stacking of $R$ triangular lattice Ga honeycomb layers (Fig. 2A) [32,33]. ErGa$_2$ exhibits an archetypical two-step metamagnetic transition for $H||c$ [34], desirable for the current study. This is due to the frustration [18] among the Er moments with strong easy-axis type anisotropy [18,35]. Below the antiferromagnetic transition at $T_N$ = 7 K [36], commensurate single-$q$ order for the up-down configuration of Er-moments is stabilized in zero field (AFM in Fig. 2A) [37]. Application of magnetic field along the easy axis induces two stepwise transitions [18,36,37]. At magnetic field of 0.8 T a 3up-1down type commensurate phase is induced. The network of upward magnetic sites forms a kagome network (Fig. 2A) [18]. This phase retains the original magnetic modulation vector ($q_{AFM}$ = ½$a^*$) as the principal modulation, but forms a triple-$q$ state with a new reciprocal lattice unit: $a^*_{kagome}$ = ½$a^*$, $b^*_{kagome}$ = ½$b^*$, and $c^*_{kagome}$ = $c^*$, where $a^*$, $b^*$, and $c^*$ are the reciprocal lattice unit of the original lattice. At low temperature, a half-magnetization plateau remains until the field-induced ferromagnetic state for $q_{FM}$ = (0, 0, 0) is induced at 2 T (Fig. 2A). Coupling between conduction electrons and Er moments is observed as a sudden drop of resistivity at $T_N$ [38] as well as in magnetoresistance anomalies across the metamagnetic transitions [39].

Figures 2B-D show magnetization and magnetotransport properties of ErGa$_2$, measured with single crystals for $H||c$. The half-magnetization plateau in the intermediate kagome spin state (Fig. 2E) and characteristic magnetic magnetoresistance are similar to those previously reported [39]. Across the metamagnetic transition, we further observe a non-monotonic Hall resistivity, which shows a positive enhancement in the intermediate kagome state. As a THE is precluded by the Ising-type magnetism, this arises from some combination of OHE and AHE. The persistence of the non-linearity at $T$ = 10 K in the paramagnetic phase (gray curve in Fig. 2D) is suggestive of two-carrier behavior with high-mobile hole-type (low density) carriers and electron-type low mobile (high density) carriers, which sharpens at lower temperatures along with magnetization and $\rho_{xx}$ (see Fig. 2C), *i.e.* an OHE.

Comparison of the temperature dependence of $\rho_{yx}/B$ of ErGa$_2$ and that for isoelectronic and isostructural LaGa$_2$ [40] (see Fig. 2F) shows that the two are of a similar value at all the temperature range (ratio < 1.4) including lack of significant discrepancy below $T_N$ for ErGa$_2$, suggesting a weakness or absence of AHE in ErGa$_2$. This is consistent with the absence of skew scattering given the longitudinal conductivity $\sigma_{xx}$ ~ $10^5$ S/cm [41], where the intrinsic mechanism rather than the extrinsic scattering mechanism becomes more relevant. Moreover, as shown in Fig. 2F, there is no significant anomaly in $\rho_{yx}/B$ at $T_N$ in contrast to the factor $\rho_{xx}^2 M/B$ (blue circle in Fig. 2F) where the intrinsic Berry curvature would manifest if present [42]. This is distinct from the reported features for other rare-earth materials [14,15,43-45] and $R$Ga$_2$ ($R$ = Ce, Sm, Gd) [40,46], where anomalous Hall contribution was suggested. We note that demagnetization correction may also affect the estimation of $R_0$ as discussed in SI Sec. S2; therein, the Hall coefficient is enhanced at low temperatures and shows a peak at $T_N$ (not explicitly argued in the previously [40,46]). The absence of AHE in ErGa$_2$ may further arise due to small de Gennes factors ($\propto (g_J - 1)^2 J(J+1)$), the square root of which roughly determines the scale of the exchange field on conduction electrons from the localized moment of rare earth ions [47,48].

To illustrate the consistency of the observed effect with the MM Hall effect, we plot $M$, $\sigma_{xx}$ ($=\rho_{xx}/(\rho_{xx}^2+\rho_{yx}^2)$), and $\sigma_{xy}$ ($=\rho_{yx}/(\rho_{xx}^2+\rho_{yx}^2)$) at $T$ = 10 K in Figs. 3A-C as a function of $B$ (= $\mu_0 H_{int}$ + $M$). We analyze the transport properties within the model introduced above. For simplicity, we introduce the mobility modulation due to spin alignment to the electron-type carriers (dominant in ErGa$_2$, which we return to below). Figures 3A-C show this analysis applied for the PM-to-Fi-FM process at $T$ = 10 K. The field dependence of $\sigma_{xx}$ and $\sigma_{xy}$ can be simulated within a conventional two-band model:

$$\sigma_{xx}(B) = \sigma_{xx}^{e} + \sigma_{xx}^{h} = \frac{en_e\mu_e(B)}{1+[\mu_e(B)B]^2} + \frac{en_h\mu_h}{1+(\mu_h B)^2},$$
$$\sigma_{xy}(B) = \sigma_{xy}^{e} + \sigma_{xy}^{h} = -\frac{en_e\mu_e(B)^2 B}{1+[\mu_e(B)B]^2} + \frac{en_h\mu_h^2 B}{1+(\mu_h B)^2}, \quad (5)$$

where $\mu_h$, $n_e$, and $n_h$ are fitting parameters. From the field dependence of $M(B)$, $\mu_e(B)$ is captured by fixing $m_e/\tau^e_0$ and $m_e/\tau^e_{mag0}$ in Eqs. (3)-(4) (the negligible contributions from the spin-fluctuations and spin-wave-scattering are further discussed in SI Sec. S3). This shows reasonable agreement with the observed response, reproducing the spin MR $\sigma_{xx}$ and nonmonotonic behavior in $\sigma_{xy}$ simultaneously.

We next extend this model to $T$ = 1.8 K within the AFM, kagome, and Fi-FM phases as shown in Figs. 3D-F, where we assume the proportionality of the field dependence of $\rho_{xx}(B)/\rho_{xx0}$ to $1/\mu_e(B)$ (see Fig. 3D) [31]. We note that at low temperatures there are interphase states (IP1 and IP2) across the metamagnetic transitions which give rise to singular peaks in $\sigma_{xx}$ (see Fig. 2E); as these are proposed to arise from specular domain wall scattering [39], we exclude these in the following analysis (gray hatched area in Figs. 3D-F, see SI Sec. S4 for a discussion on the IP states). We also note that multi-step metamagnetism is observed (see SI Sec. S2) suggestive of more complex intermediate spin configurations in the IP1 and IP2 regions (this is beyond the scope of our simplified model).

Figures 3G-H summarize the transport parameters obtained by performing the analysis using the ansatz $\rho_{xx}(B)/\rho_{xx0} \propto 1/\mu_e(B)$ at various temperatures. For $T$ = 10 and 8 K (> $T_N$), the parameter set is consistent with those for Eq. (4). We also confirmed the absence of significant anomaly in $\mu_e$(9 T), $n_e$, $n_h$, and $\mu_h$ at $T_N$, while $\mu_e$(0 T) is enhanced below $T_N$ due to the suppression of the spin-disorder scattering. $n_h$ shows a slight temperature dependence (Fig. 3H), which might not be an intrinsic effect as the fitting quality does not change by fixing $n_e$ and $n_h$ constants. Figure 3I shows the zero field transport properties reproduced from the fitting parameters; the two-band Hall coefficient ($R_0$ in Eq. (2)) and resistivity $\rho_{xx,0T}$ (= $\sigma_{xx}(0\,T)^{-1}$ in Eq. (5)) are reasonably consistent with the direct observation. The numbers of carrier per unit volume are 1.2 and 3×10$^{-4}$, for electron and hole, respectively. The large electron concentration is consistent with the negative slope of $\rho_{yx}$ at high fields, corresponding to 1.5×10$^{22}$ cm$^{-3}$ (Fig. 2D). Two orders of magnitude higher mobility of the hole pocket are necessary to reproduce the positive Hall coefficient at zero field (Figs. 2F, 3I). We note that these values reflect an oversimplification of the electronic structure of this system as a two-band model, and are not directly connected to the volume of Fermi surfaces for the low-field Hall effect [49-51], which is affected by the local curvatures (see below).

To evaluate the relevance for this model in the present intermetallic setting, we calculate the band structure for the non-magnetic analog LaGa$_2$ (see Fig. 4A). Ellipsoidal hole pockets (FS1 and FS2, see Fig. 4B) at the A point on the BZ edge are ascribed to the Ga-4p band, while the large Fermi surface (FS3, see Fig. 4C) is comprised mainly of La-5d orbitals, consistent with the previous studies [52, 53]. These two bands correspond to those in the schematic model in Fig. 1A. The Hall coefficient for each Fermi surface is determined by the sign of curvature $\gamma$ ($\propto -\sigma_{xy}$) [49,50]. We map $\gamma$ onto each Fermi surface as shown in Figs. 4D-E. For the hole bands (FS1 and 2), it is evident that their contributions to the Hall effect are $\sigma^h_{xy}$ > 0. The FS3 has both characters showing a sign change for $\gamma$ from negative near $k_z = \pi/c$ (A point) to positive across around $k_z = \pi/2c$ to 0 (Fig. 4E), where $c$ is the lattice constant. The representative electron-type carrier orbits at $k_z \sim \pi/8c$ and $k_z \sim$ 0, where $c$ is the lattice constant, are depicted in Figs. 4F-G, respectively, which confirms the presence of an electron type orbit deflecting counter-clock-wise (see SI Sec. S5) [51]. We note that a three-band treatment (a hole pocket for p-band, an electron and a hole pockets with comparable concentration and mobility for d-band) would be more realistic on the basis of the band calculations. This can be effectively treated by a two-band model by representing the d-band pockets as a large electron pocket and leaving a hole pocket for the high-mobile p-band (see SI Sec. S6). These results support the application of the two-band model with the asymmetric spin-charge coupling as an ansatz to analyze the magnetotransport properties of $R$Ga$_2$. We also note that this argument can be applied to the magnetic ErGa$_2$ on the basis of the rigid band approximation from the nonmagnetic analogue LaGa$_2$, where the 4f electrons are expected to be well localized (Fig. 1A) and only weakly hybridize with the electrons near the Fermi energy. This is consistent with previous quantum oscillation experiments in various $R$Ga$_2$ ($R$ = Ce, Pr, Sm, Gd) [46, 54-57]. In these studies, the quantum oscillation branches associated with the FS1, 2, and 3 are identified almost unchanged from the nonmagnetic LaGa$_2$. Further, in CeGa$_2$, the exchange splitting on the FS1 and FS2 has been observed to be smaller than that of FS3 [54] validating the above simplification for the mobility modulation only on the electron-type carriers in ErGa$_2$.

**Discussion**

We note that the modulation of the ordinary Hall effect has also been considered in dilute magnetic alloys, therein known as the "spin effect" [58]. In particular, it has been established that a variation of $R_0$ arises due to the difference of the field-induced change of scattering rate of conduction electrons with up and down spins. The distinction from the current model is that the former always leads to the enhancement of $R_0$, but the latter can induce both enhancement and reduction depending on the details of the electronic and magnetic subsystems. In the current analysis, we consider the mobility is field-dependent in the magnetization process. It is known that the mobility can be field-dependent even without the magnetic moments as observed in diluted (semi)metals [59, 60] when the disorder potential smoothly varies compared to the cyclotron radius [61]. This effect is expected to be masked in the present metallic system owing to the short Thomas-Fermi screening length (~Å). Magnetic semiconductors/semimetals with reduced carriers are a class of materials where this effect potentially interferes with the MM Hall effect. Beyond the mobility and effective mass changes discussed above, we also note that there are possible contributions to the field-dependence of carrier density across metamagnetic transitions. This effect could be significant in semimetals with low carrier concentration as discussed previously in Eu compounds [62]. ErGa$_2$, on the other hand, is a metal with large carrier number and not a heavy-fermion system and is anticipated to be insensitive to changes of magnetic $q$-vector as it is commensurate to the lattice. The valley dependent carrier emptying [63] as well as the magnetic breakdown [64] is irrelevant in ErGa$_2$ (and is instead important for the quantum limit in semimetals).

In conclusion, we have observed a metamagnetic multiband Hall effect in the Ising-antiferromagnet ErGa$_2$ and demonstrated its consistency with a realistic two-band model. The orbital character in the band structure is further consistent with the two-carrier feature with different scattering cross section to magnetic moments. The band-induced Hall effect discussed in our work in principle can also be present in intermetallic metamagnetic systems with topological spin textures. For the Hall anomaly in such systems, this shows rather than a straightforward assignment to a topological Hall effect via the ansatz Eq. (1), that instead it can be captured by a multiband response of the charge carriers. This finding provides a clear example where the ansatz Eq. (1) fails to capture the origin of the Hall response and a simple framework to understand the magnetotransport properties of magnetic intermetallics, which would be useful for more precise evaluation of the emergent field therein.

Note added: Recently, we became aware of a related work by Y. Ōnuki *et al.* [65] on transport properties of $R$Ga$_2$ ($R$ = rare earth) and structurally related systems.

**Materials and Methods**

Single crystals of ErGa$_2$ were grown via the Pb self-flux method following Ref. [66]. The starting materials Er, Ga, and Pb were mixed in the molar ratio Er : Ga : Pb = 1 : 2 : 10. They were loaded into a 2-mL alumina crucible and sealed in an evacuated quartz tube. The growth ampoule was heated to 1000 °C and slowly cooled to 400 °C at a rate of 1 °C/min. To increase the size of crystals, the furnace temperature was increased to 850 °C in 2 hours and cooled down to 400 °C at 1 °C/min; this process was subsequently repeated between 700 °C to 400 °C. The single crystals were separated by decanting the flux in a centrifuge. Finally, crystals were annealed in an evacuated quartz tube at 600 °C for 6 days. The typical size of the crystals is 1×1×0.5 mm$^3$ (for $a \times b \times c$). The quality of the crystal was checked by the single crystal x-ray diffraction using the synchrotron light source at SPring-8, Japan (see SI Sec. S7).

Electrical transport measurements were performed by a conventional five probe method with an AC excitation current of 1 mA at typical frequency near 15 Hz. The transport response in low temperature and a magnetic field was measured using a commercial superconducting magnets and cryostat. The obtained longitudinal and transverse signals were field symmetrized and antisymmetrized to correct for a contact misalignment, respectively. Magnetization measurements were performed using a commercial superconducting quantum interference device (SQUID) magnetometer.

The electronic band structures of LaGa$_2$ were calculated by the density functional theory (DFT) code using the Vienna ab initio simulation package (VASP) (see SI Sec. S5). The electronic band structures and projections were further computed on a finer 100×100×90 $k$-mesh grid to create interpolated Fermi surfaces near the Fermi level for evaluating band characteristics and curvatures.

**Acknowledgments**

We thank Y. Nakamura, and H. Sawa for supporting synchrotron XRD experiments. The synchrotron radiation experiments were performed at SPring-8 with the approval of the Japan Synchrotron Radiation Research Institute (JASRI) (Proposal No. 2023A1882). **Funding:** This research is funded in part by the Gordon and Betty Moore Foundation EPiQS Initiative through Grants GBMF9070 to J.G.C. (material synthesis, DFT calculations), NSF grant DMR-1554891 (material design), ONR Grant N00014-21-1-2591 (instrumentation development), and AFOSR grant FA9550-22-1-0432 (advanced characterization). T.K. acknowledges the support by the Yamada Science Foundation Fellowship for Research Abroad and JSPS Overseas Research Fellowships. L.Y. acknowledges support by the STC Center for Integrated Quantum Materials, NSF grant number DMR-1231319, the Heising-Simons Physics Research Fellow Program, and the Tsinghua Education Foundation.

**References**

1. C. M. Hurd, The Hall effect in metals and alloys, Plenum, New York (1972).

2. Y. Tokura, and N. Kanazawa, Magnetic skyrmion materials, Chem. Rev. **121**, 2857 (2020).

3. X. G. Wen, F. Wilczek, and A. Zee, Chiral spin states and superconductivity, Phys. Rev. B **39**, 11413 (1989).

4. J. Ye et al., Berry phase theory of the anomalous Hall effect: application to colossal magnetoresistance manganites, Phys. Rev. Lett. **83**, 3737 (1999).

5. P. Bruno, V. K. Dugaev, and M. Taillefumier, Topological Hall effect and Berry phase in magnetic nanostructures, Phys. Rev. Lett. **93**, 096806 (2004).

6. S. D. Yi et al., Skyrmions and anomalous Hall effect in a Dzyaloshinskii-Moriya spiral magnet, Phys. Rev. B **80**, 054416 (2009).

7. Y. Taguchi et al., Spin chirality, Berry phase, and anomalous Hall effect in a frustrated ferromagnet, Science **291**, 2573 (2001).

8. A. Neubauer et al., Topological Hall effect in the *A* phase of MnSi, Phys. Rev. Lett. **102**, 186602 (2009).

9. M. Lee et al., Unusual Hall effect anomaly in MnSi under pressure, Phys. Rev. Lett. **102**, 186601 (2009).

10. R. Ritz et al., Giant generic topological Hall resistivity of MnSi under pressure, Phys. Rev. B **87**, 134424 (2013).

11. C. Franz et al., Real-space and reciprocal-space Berry phases in the Hall effect of $Mn_{1-x}Fe_xSi$, Phys. Rev. Lett. **112**, 186601 (2014).

12. T. Kurumaji et al., Skyrmion lattice with a giant topological Hall effect in a frustrated triangular-lattice magnet, Science **365** 914 (2019).

13. W.-L. Lee et al., Dissipationless anomalous Hall current in the ferromagnetic spinel $CuCr_2Se_{4-x}Br_x$, Science **303**, 1647 (2004).

14. J. J. Rhyne, Anomalous Hall effect in single-crystal dysprosium, Phys. Rev. **172**, 523 (1968).

15. J. J. Rhyne, Anomalous and ordinary Hall effect in terbium, J. Appl. Phys. **40**, 1001 (1969).

16. A. Gerber, Interpretation of experimental evidence of the topological Hall effect, Phys. Rev. B **98**, 214440 (2018).

17. G. Kimbell et al., Challenges in identifying chiral spin textures via the topological Hall effect, Commun. Mater. **3**, 19 (2022).

18. M. Doukoure, and D. Gignoux, Metamagnetism in $ErGa_2$ studied on a single crystal, J. Magn. Magn. Mater. **30**, 111 (1982).

19. E. H. Sondheimer, and A. H. Wilson, the theory of the magneto-resistance effects in metals, Proc. Roy. Soc. A, **190**, 435 (1947).

20. V. F. Ghantmakher, and Y. B. Levinson, Carrier scattering in metals and semiconductors, North-Holland, Amsterdam (1987).

21. J. Kuneš, and W. E. Pickett, Kondo and anti-Kondo coupling to local moments in $EuB_6$, Phys. Rev. B **69**, 165111 (2004).

22. H.-S. Li, Y. P. Li, and J. M. D. Coey, R-T and R-R exchange interactions in the rareearth (R)-transition-metal (T) intermetallics: an evaluation from relativistic atomic calculations, J. Phys.: Condens. Matter **3**, 7277 (1991).

23. T. Nomoto, T. Koretsune, and R. Arita, Formation mechanism of the helical *Q* structure in Gd-based skyrmion materials, Phys. Rev. Lett. **125**, 117204 (2020).

24. K. Yosida, Anomalous electrical resistivity and magnetoresistance due to an *s-d* interaction in Cu-Mn alloys, Phys. Rev. **107**, 396 (1957).

25. K. Kubo, and N. Ohata, A quantum theory of double exchange. I, J. Phys. Soc. Jpn. **33**, 21 (1972).

26. A. Urushibara, Y. Morimoto, T. Arima, A. Asamitsu, G. Kido, and Y. Tokura, Insulator-metal transition and giant magnetoresistance in $La_{1-x}Sr_xMnO_3$, Phys. Rev. B **51**, 14103 (1995).

27. T. Kasuya, Electrical resistance of ferromagnetic metals, Prog. Theo. Phys. **16**, 58 (1956).

28. P. G. de Gennes, and J. Friedel, Anomalies de resistivite dans certains metaux, magnetiques, J. Phys. Chem. Solids **4**, 71 (1958).

29. T. van Peski-Tinbergen, and A. J. Dekker, Spin-dependent scattering and resistivity of magnetic metals and alloys, Physica **29**, 917 (1963).

30. S. M. Thomas et al., Hall effect anomaly and low-temperature metamagnetism in the Kondo compound $CeAgBi_2$, Phys. Rev. B **93**, 075149 (2016).

31. L. Ye, T. Suzuki, and J. G. Checkelsky, electronic transport on the Shastry-Sutherland lattice in Ising-type rare-earth tetraborides, Phys. Rev. B **95**, 174405 (2017).

32. A. Raman, On $AlB_2$ type phases, Zeitschrift für Metallkunde, 58, 179 (1967).

33. A. Fedorchuk, and Y. Grin, Crystal structure and chemical bonding in gallides of rare-earth metals, Handbook on the Physics and Chemistry of Rare Earths, 53, 81 (2018).

34. T.-H. Tsai, and D. J. Sellmyer, Magnetic ordering and exchange interactions in the rare-earth gallium compounds $RGa_2$, Phy. Rev. B **20**, 4577 (1979).

35. M. Diviš, M. Richter, J. Forstreuter, K. Koepernik, and H. Eschrig, Crystal-field and magnetism of $RGa_2$ ($R$ = Ce, Er) compounds derived from density-functional calculations, J. Magn. Magn. Mater. 176, L81 (1997).

36. T. H. Tsai, J. A. Gerber, J. W. Weymouth, and D. J. Sellmyer, Magnetic properties of the rare-earth intermetallics $RGa_2$, J. Appl. Phys. 49, 1507 (1978).

37. H. Asmat, and D. Gignoux, Magnetic structures of the intermetallic rare-earth compounds $RGa_2$, Inst. Phys. Conf. Ser. **37**, 286 (1978).

38. J. A. Blanco, D. Gignoux, J. C. Gómez Sal, J. Rodríguez Fdez, and D. Schmitt, Magnetic contribution to the electrical resistivity in $RGa_2$ compounds ($R$ = rare earth), J. Magn. Magn. Mater. 104-107, 1285 (1992).

39. N. V. Baranov, P. E. Markin, A. I. Kozlov, and E. V. Sinitsyn, Effect of magnetic interphase boundaries on the electrical resistivity in metallic metamagnets, J. Alloys Comp. 200, 43 (1993).

40. H. Henmi, Y. Aoki, T. Fukuhara, I. Sakamoto, and H. Sato, Transport properties of $RGa_2$ ($R$ = La, Ce and Sm), Physica B 186-188, 655 (1993).

41. S. Onoda, N. Sugimoto, and N. Nagaosa, Quantum transport theory of anomalous electric, thermoelectric, and thermal Hall effects in ferromagnets, Phys. Rev. B **77**, 165103 (2008).

42. M. Lee, Y. Onose, Y. Tokura, and N. P. Ong, Hidden constant in the anomalous Hall effect of high-purity magnet MnSi, Phys. Rev. B **75**, 172403 (2007).

43. F. E. Maranzana, Contributions to the theory of the anomalous Hall effect in ferro- and antiferromagnetic materials, Phys. Rev. **160**, 421 (1967).

44. B. Giovannini, theory of the anomalous Hall effect in the rare earths, Phys. Lett. A **27**, 381 (1971).

45. T. Hiraoka, Antiferromagnetic skew scattering Hall effect in NdZn, J. Phys. Soc. Jpn. **55**, 4417 (1986).

46. S. Ohara, I. Sakamoto, Y. Aoki, H. Sato, I. Oguro, T. Sasaki, G. Kido, S. Maruno, Magnetic properties and de Haas-van Alphen effect of $GdGa_2$, Physica B **223&224** 379, (1996).

47. S. H. Liu, Exchange interaction between conduction electrons and magnetic shell electrons in rare-earth metals, Phys. Rev. **121**, 451 (1961).

48. J. Kondo, Anomalous Hall effect and magnetoresistance of ferromagnetic metals, Prog. Theor. Phys. **27**, 772 (1962).

49. F. D. M. Haldane, A geometrical description of the "classical" weak-field Hall effect in metals, arXiv:0504227 (2005).

50. O. Narikiyo, Fermi-surface curvature and Hall conductivity in metals, J. Phys. Soc. Jpn. 89, 124701 (2020).

51. N. P. Ong, Geometric interpretation of the weak-field Hall conductivity in two-dimensional metals with arbitrary Fermi surface, Phys. Rev. B **43**, 193 (1991).

52. H. Harima, and A. Yanase, Electronic structure and Fermi surface of $LaGa_2$, J. Phys. Soc. Jpn. 60, 2718 (1991).

53. M. Sahakyan, V. H. Tran, Density functional theory study of electronic structure and optical properties of $YGa_2$, Comp. Mater. Sci. 184, 109898 (2020).

54. I. Umehara, N. Nagai, and Y. Onuki, Magnetoresistance and de Haas-van Alphen effect in $CeGa_2$, J. Phys. Soc. Jpn. **60**, 1464 (1991).

55. I. Umehara, N. Nagai, A. Fukuda, K. Satoh, Y. Fujimaki, and Y. Onuki, Fermi surface and cyclotron mass in $CeGa_2$, J. Magn. Magn. Mater. **104-107**, 1407 (1992).

56. I. Sakamoto, T. Miura, H. Sato, T. Miyamoto, I. Shiozaki, I. Oguro, and S. Maruno, Fermi surface and magnetic properties of antiferromagnetic $SmGa_2$, J. Magn. Magn. Mater. **108**, 125 (1992).

57. I. Sakamoto, Y. Isokane, H. Sato, K. Maezawa, G. Kido, I. Oguro, and S. Maruno, De Haas-van Alphen effect and magnetic properties of antiferromagnetic $SmGa_2$ and $PrGa_2$, Physica B **194-196**, 1175 (1994).

58. A. Fert et al., J. Magn. Magn. Mater. **24**, 231 (1981).

59. B. Fauqué *et al.*, Magnetoresistance of semimetals: the case of antimony, Phys. Rev. Mater. **2**, 114201 (2018).

60. C. Collignon *et al.*, Quasi-isotropic orbital magnetoresistance in lightly doped $SrTiO_3$, Phys. Rev. Mater. **5**, 065002 (2021).

61. J. C. W. Song, G. Rafael, and P. A. Lee, Linear magnetoresistance in metals: guiding center diffusion in a smooth random potential, Phys. Rev. B **92**, 180204(R) (2015).

62. H. W. Meul et al., Phys. Rev. B **26**, 6431 (1982).

63. Z. Zhu, J. Wang, H. Zuo, B. Fauqué, R. D. McDonald, Y. Fuseya, and K. Behnia, Emptying Dirac valleys in bismuth using high magnetic fields, **8**, 15297 (2017).

64. L. M. Falicov, and P. R. Sievert, Magnetoresistance and magnetic breakdown, Phys. Rev. Lett. **12**, 558 (1964).

65. Y. Ōnuki et al., Anomalous Hall effect in rare earth antiferromagnets with the hexagonal structures, New Physics: Sae Mulli **73**, 1054 (2023).

66. R. D. dos Reis et al., Anisotropic magnetocaloric effect in $ErGa_2$ and $HoGa_2$ single-crystals, J. Alloys Comp. **582**, 461 (2014).

**Figures and Tables**

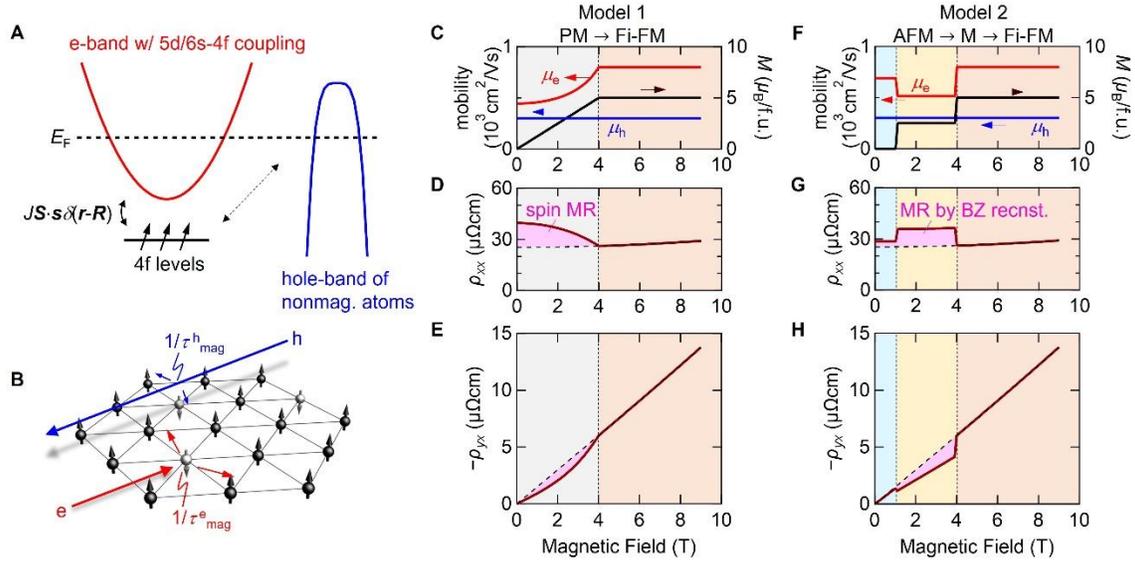

**Figure 1. A model for the metamagnetic multiband Hall effect**
**A** Schematic electronic structure for two-band model. The red band is electron-like and made from cation 5d/6s orbitals, which couples to the underlying 4f electrons through $J\mathbf{S}\cdot\mathbf{s}\delta(\mathbf{r}-\mathbf{R})$. The blue band is hole-like and composed of s/p orbitals of nonmagnetic electronegative atoms (anions). **B** Schematic charge transport in the magnetic lattice with spin disorder. **C**-**E** Calculated field dependence of **C** magnetization ($M$) and electron ($\mu_e$) /hole ($\mu_h$) mobilities, **D** resistivity ($\rho_{xx}$), **E** Hall resistivity ($\rho_{yx}$) for Model 1, where the term spin MR denotes the low-field negative MR highlighted by pink that is due to the alignment of the disordered spins. **F-H** Corresponding plot for Model 2, where the MR is due to the effective-mass renormalization by the BZ reconstruction across metamagnetic transitions. The carrier densities are $n_e = 2.5\times10^{20}$ cm$^{-3}$ and $n_h = 1.6\times10^{20}$ cm$^{-3}$, which are selected to illustrate the model. The dashed line in **D-H** represents fixed $\mu_e$.

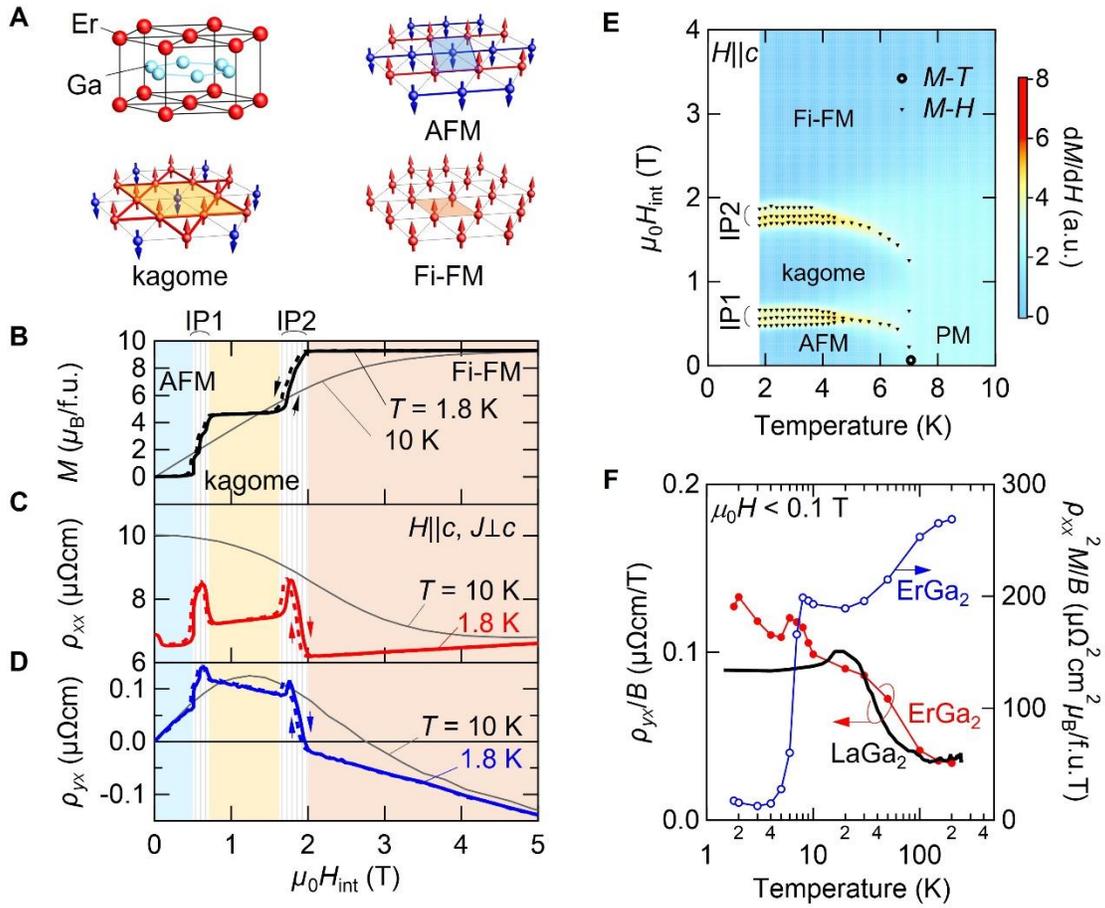

**Figure 2. Metamagnetic Multiband Hall response in metamagnetic ErGa$_2$**
**A** Crystal structure and schematic magnetic structure of the Er triangular-lattice for each magnetic phase for $H||c$. AFM: commensurate single-$q$ antiferromagnetic, kagome: commensurate triple-$q$ state, Fi-FM: field-induced ferromagnetic phases, respectively. Hatched area highlights the magnetic unit cell. **B** $M$, **C** $\rho_{xx}$, and **D** $\rho_{yx}$ for $H||c$ as a function of the internal magnetic field $\mu_0 H_{int} = \mu_0 H_{ext} - NM$. Gray-hatched regions are the interphase states (IP1 and IP2, see SI Sec. S2) discussed in Ref. [39]. The excitation current ($J$) is applied in the $ab$-plane. The thick solid (dashed) curve is for field-increase (decrease) scan at $T = 1.8$ K, and the thin gray curve is for $T = 10$ K. **E** Magnetic phase diagram in $H||c$. **F** Red: Temperature dependence of Hall coefficient ($\rho_{yx}/B$) of ErGa$_2$, where $B = \mu_0 H_{int} + M$; blue: scaling factor for supposed intrinsic anomalous Hall effect ($\rho_{xx}^2 M/B$). The black curve is the Hall coefficient of LaGa$_2$ reproduced from Ref. [40].

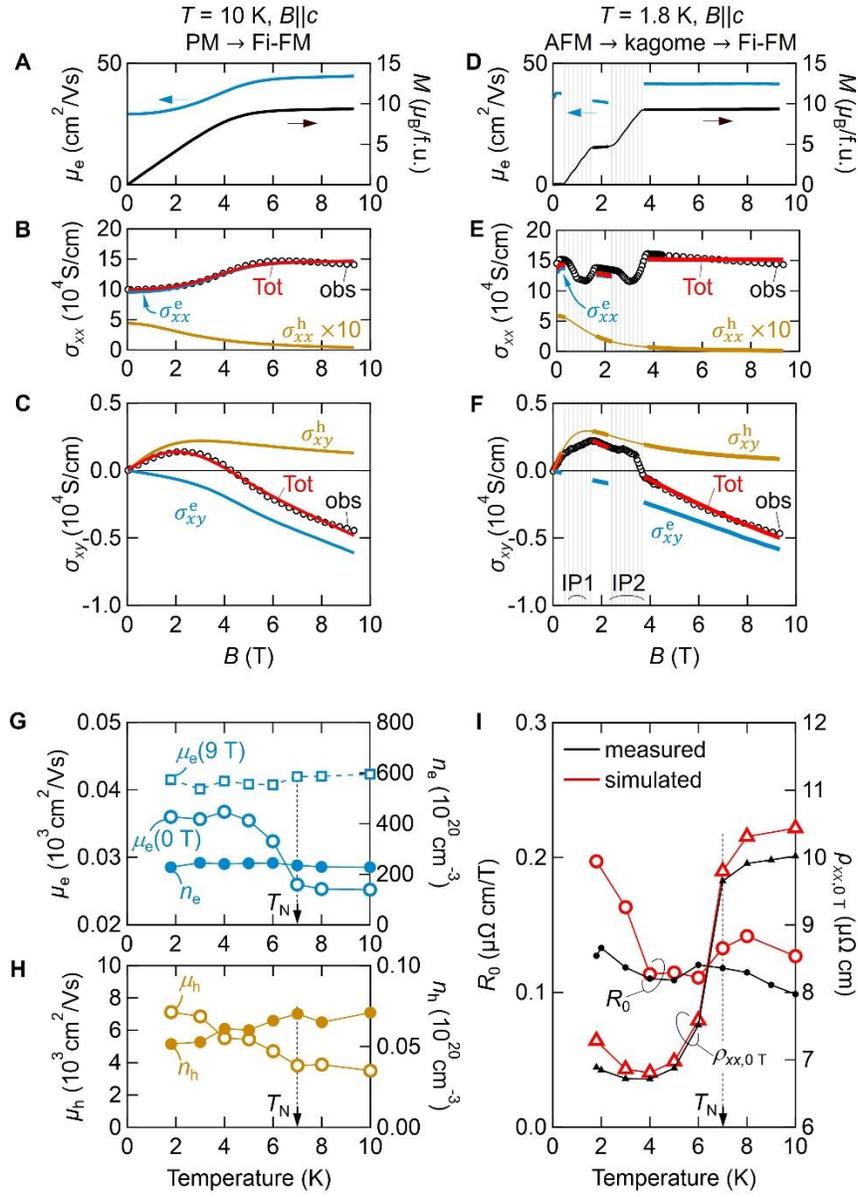

**Figure 3. Simulations of the conductivity tensors of ErGa$_2$**
**A** (**D**) Field dependence of the simulated $\mu_e$ (for Eq. (5)) and $M$, **B** (**E**) conductivity ($\sigma_{xx}$), and Hall conductivity ($\sigma_{xy}$) at $T$ = 10 (1.8) K. In **B**-**C** (**E**-**F**), black open circles are observation (obs), and red (Tot), cyan ($\sigma_{xx}^e$ and $\sigma_{xy}^e$), and yellow ($\sigma_{xx}^h$ and $\sigma_{xy}^h$) curves are for the total (Eq. (5)), electron, and hole contributions, respectively. For $T$ = 1.8 K, the hatched area is for the IP1 and IP2 states. **G** Temperature dependence of the simulated electron mobility $\mu_e(B)$ at $B$ = 0 T (open circle) and at 9 T (open square), and electron charge density $n_e$ (closed circle). **H** Temperature dependence of fitting parameters for holes: mobilities $\mu_h$ and charge density $n_h$. Transition temperature ($T_N$) at zero field is also denoted. **I** Temperature dependence of $R_0$ (Eq. (2)) and zero-field $\rho_{xx}$. Closed (open) markers are obtained (simulated) from the direct measurement (fitting parameters).

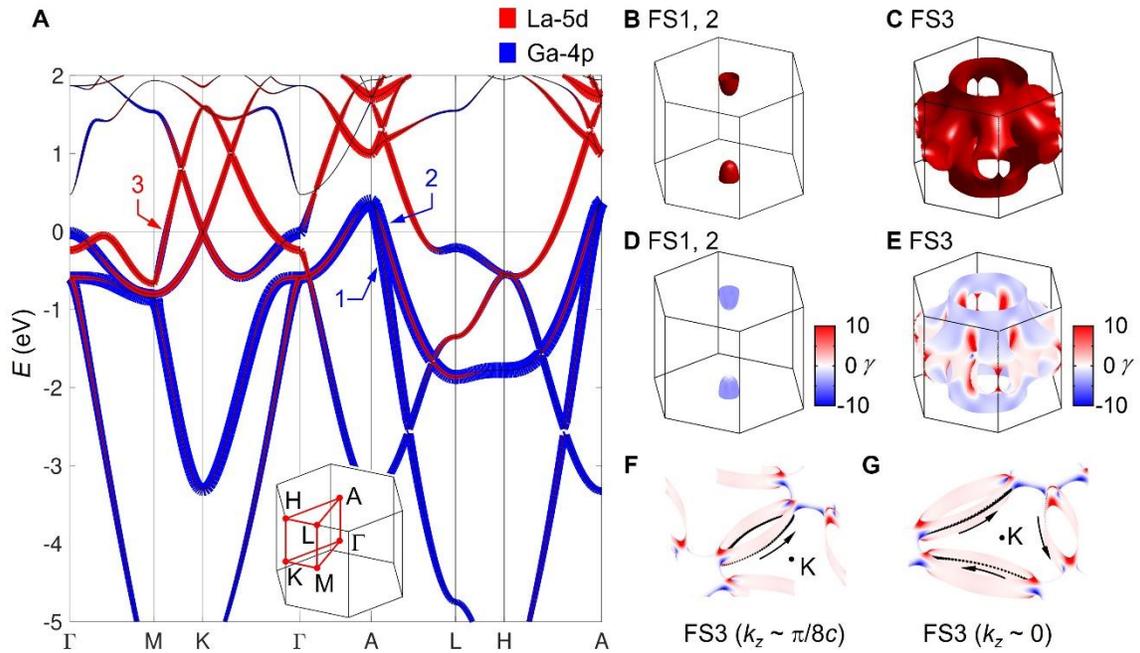

**Figure 4. Fermi surfaces and electron/hole orbits in LaGa$_2$.**
**A** DFT calculation of the band structure of LaGa$_2$. Color scale represents the orbital projections to La-5d (red) and Ga-4p (blue). Band 1, 2, and 3 are related to the FS1, 2, 3, respectively. Inset shows the first BZ. **B**-**C** Fermi surfaces of hole bands FS1, FS2, and FS3. **D**-**E** Color map of the curvature-field $\gamma(\mathbf{s})$ at the Fermi surface piece $\mathbf{s}$ (see SI Sec. S5). Red (blue) contributes to electron (hole) type Hall effect. **F**-**G** Representative electron orbits in the reciprocal space for the FS3.

Supplemental material for

**Metamagnetic multiband Hall effect in Ising antiferromagnet ErGa$_2$**

T. Kurumaji *et al*.

**S1. Variation of the models with different types of carrier scattering.**

**S2. Demagnetization correction to magnetic and transport properties in ErGa$_2$.**

**S3. Estimation of spin-disorder scattering in ErGa$_2$.**

**S4. Interpolation of the simulations into the IP states.**

**S5. Band structure calculations of LaGa$_2$.**

**S6. Simulation curves by using different parameters.**

**S7. Single crystal x-ray diffraction for ErGa$_2$.**

**S1. Variation of the models with different types of carrier scattering.**

We consider the models for the MM Hall effect in the main text, where only the electron bands are supposed to interact with the magnetic subsystem through the intraatomic f-d coupling. This assumption is relevant for the analysis on ErGa$_2$. Here, we consider the interatomic coupling between the 4f moments and the anionic 4p bands, which is enabled by the hybridization of these bands through the mechanism proposed in Ref. [S1]. Figures S1A-C is for the MM Hall effect when the hole mobility modulation alone is significant (Model 3). For simplicity, we consider the paramagnetic effect, which can be straightforwardly extended to the metamagnetic system. The negative MR (spin MR, see Fig. S1B) is induced by the modulation of the mobility (Fig. S1A), and the hump appears in $\rho_{yx}$ (Fig. S1C). This is qualitatively similar to Model 1 (Fig. 1E), while the sign of Hall anomaly is opposite. This is reasonable as the mobility is modulated for different band with opposite sign of carriers.

As another variation of the model, we introduce the Model 4, where the mobility modulation is considered for both bands. The magnitude of the modulation for the electron band is fixed and that of the hole band is varied (see Fig. S1D). Correspondingly, the spin MR is seen in all cases, and the MM Hall anomaly shows a cross over from the dip to hump as the effect of the hole band modulation is relevant. These results suggest that the sign of the MM Hall effect gives an insight on which band is more relevantly modified by the magnetic interactions.

**S2. Demagnetization correction to magnetic and transport properties in ErGa$_2$.**

We performed demagnetization corrections for the magnetization and magnetotransport properties of ErGa$_2$. Figure S2A shows the relationship among three related fields: the external field $\mu_0 H_{\text{ext}}$, the internal magnetic field $\mu_0 H_{\text{int}} = \mu_0 H_{\text{ext}} - N_d M$, and the magnetic induction $B = \mu_0 H_{\text{int}} + M$. As the demagnetization field $N_d M$ is opposite to $\mu_0 H_{\text{ext}}$, the internal field $\mu_0 H_{\text{int}}$ (red line) is reduced relative to $\mu_0 H_{\text{ext}}$. The field dependence of $M$ shows a stepwise transition when plotted as a function of $\mu_0 H_{\text{int}}$ (Fig. S2B), characteristic of metamagnetism. The Hall coefficient $R_0$ is obtained by dividing $\rho_{yx}$ by $B$ as shown in Fig. S2C [S2, S3]. We also show $\rho_{yx}/\mu_0 H_{\text{int}}$, which can be compared to previous reports in $R$Ga$_2$ (R = Ce, Sm, Gd) [40,46]. We plot the field dependence of magnetization and magnetotransport properties at various temperatures in Figs. S2D-F as a function of $\mu_0 H_{\text{int}}$.

We note that the absence of the hysteresis in magnetization in our sample, in contrast to the data in Ref. [18]. This is potentially due to the better quality with less domain-wall pinning effect. In the intermediate region, we discern a fine structure in the magnetization jumps. To more clearly see the metamagnetic transition for $H\|c$, we plot the field derivative ($dM/dH_{ext}$) of $M$ at various temperatures (Fig. S3A). We observe that $dM/dH_{ext}$ associated with the metamagnetic transitions has a triple peak structure below $T = 4$ K. We refer to these newly-found regions as the interphase states IP1 and IP2 as illustrated in the $H$-$T$ phase diagrams in Fig. 2E. Figures S3B-C show that each IP state can be further resolved into two (*e.g.* IP1a and IP1b) suggesting structure reminiscent of the devil's staircase in frustrated Ising systems [S4].

Such intermediate regions are known to enhance the magnetoresistivity [39] due to specular scattering at domain walls between different magnetic states. We plot the $B$-field dependence of $M$, $dM/dH$, $\sigma_{xx}$, and $\sigma_{xy}$ in Figs. S4A-D. A dip in $\sigma_{xx}$ (Fig. S4C) and a slope change in $\sigma_{xy}$ are correlated with the $dM/dH_{ext}$ peak. The scattering mechanism is highly dependent on the multidomain structure of the intermediate regions, and beyond the two-band model discussed in the main text. See also additional discussion in Sec. S4.

## S3. Estimation of spin-disorder scattering in ErGa$_2$.

In the main text, we hypothesize that the carrier scattering due to spin-fluctuations and spin-waves is negligible at $T < 10$ K due to the large crystal electric field splitting of the Er$^{3+}$ moments. This can be justified from the temperature dependence of $\rho_{xx}$ at zero field as follows.

The electron band dominates the resistivity of ErGa$_2$ ($\sigma_{xx}^e \gg \sigma_{xx}^h$); the temperature dependence of the resistivity can be simulated by the summation of the following three terms: $\rho_{xx}(T) = \rho_0 + \rho_{ph}(T) + \rho_{mag}(T)$, where $\rho_{mag}(T) = \rho_{sd}(1 - m(T)^2/m_0^2)$. $\rho_{sd}$ is the constant representing the spin-electron coupling [20, 27], corresponding to $m_e/e^2 n_e \tau_{mag0}$ (a composite parameter used in Eqs. (3)-(5)). $m(T)/m_0$ is proportional to the order parameter in the AFM state at zero field. In an Ising antiferromagnet, where the spin fluctuation is negligible, the longitudinal magnetic susceptibility is expressed within the mean-field model [S5] as

$$\chi_{\|c} = \frac{C_{\text{Ising}}(1 - m^2/m_0^2)}{T + T_N(1 - m^2/m_0^2)} + \chi_{\text{v.v.}}, \quad (S1)$$

where $C_{\text{Ising}} = N_A g_J^2 J^2 \mu_B^2/k_B$, $T_N = 2zJ_{zz}/k_B$, $m = N_A g_J \mu_B (\langle J_z \rangle_A - \langle J_z \rangle_B)/2$, $\chi_{\text{v.v.}}$ is the van Vleck susceptibility. From the temperature dependence of $M/H_{\text{int}}$ along the $c$ axis ($= \chi_{\|c}$), $m(T)/m_0$ can be obtained by

$$\frac{m(T)}{m_0} = \sqrt{1 - \frac{(\chi_{\|c} - \chi_{\text{v.v.}})T}{C_{\text{Ising}} - T_N(\chi_{\|c} - \chi_{\text{v.v.}})}}. \quad (S2)$$

Accordingly, the temperature dependence of the magnetic susceptibility $\rho_{mag}(T)$ can be simulated by the magnetic susceptiblity $\chi_{\|c}(T)$.

Figures S5A-B show the temperature dependence of anisotropic magnetic susceptibility ($M/H$ for $H\|c$ and $H\perp c$) and zero-field resistivity. Using Eq. (S2), $m(T)/m_0$ is shown in Fig. S5C. We set $C_{\text{Ising}} = 28$ emu K/mol, $\chi_{\text{v.v.}} = 0.1$ emu/mol,

and $T_N = 7.0$ K. $C_{Ising}$ is selected to maintain the positive value in the square root of Eq. (S2), and is close to the ideal value 30.4 emu K/mol for $J = 15/2$. Figure S5D compares the observed and simulated $\rho_{xx}$. $\rho_{sd}$ is set 3.18 µΩ/cm to connect the two plots at $T > T_N$, which is consistent with $\rho_{sd} = m_e/e^2 n_e \tau_{mag0} = 3.69$ µΩ/cm (red curve in Fig. 3B). The nonmagnetic contribution $\rho_0 + \rho_{ph} = a_0 + a_2 T^2 + a_5 T^5$ is determined by $\rho_{xx}$ in the PM state. The agreement of the two curves is reasonable, confirming the absence of the fluctuation-driven spin-scattering in ErGa$_2$ and that the magnetic resistivity is determined by the magnitude of disorder as $1 - m^2/m_0^2$.

On the basis of the above argument, we can exclude the AHE as the origin of the field-induced-modulation of $\rho_{yx}$ at 10 K (Fig. 2D) generated from the skew scattering due to fluctuating scalar spin-chirality (recently proposed in Ref. [S6]). While in general difficult to distinguish from the OHE [S7], here the strong easy-axis anisotropy (anisotropy gap ~ 6.5 meV > $k_B T$ = 0.86 meV [35]) of Er moments strongly suppresses scalar spin chirality on the triangular lattice.

**S4. Interpolation of the simulations into the IP states.**

In the main text, the IP state regions are excluded from the analysis. This is because the IP states are expected to be an inhomogeneous multidomain state among different magnetic structures, i.e., AFM and kagome states for IP1 and kagome and Fi-FM states for IP2 [39]. This in principle impedes the application of the two-band model as the two magnetic states are supposed to be represented by different carrier parameters. To see this effect, we interpolate the field-dependence of $\mu_e$ to the IP states as shown in Fig. S6A, which reflects the sharp peak in $\rho_{xx}$ (Fig. 2C). Using the same parameters for Figs. 3D-F, $\sigma_{xx}$ and $\sigma_{xy}$ in the IP states are simulated as shown in Figs. S6B-C. Agreement between the observations and the simulation curves is good apart from a small discrepancy in the $\sigma_{xy}$ in the IP1 state (Fig. S6C). These results suggest that the effect of the specular domain-wall scattering does not significantly reduce the relevance of the two-band model treatment in this system.

**S5. Band structure calculations of LaGa$_2$.**

The electronic band structure was calculated by the density functional theory (DFT) code using the Vienna ab initio simulation package (VASP) [S8,S9]. The ground state for LaGa$_2$ was converged with a 350 eV plane wave basis cut-off and a Γ-centered 13×13×11 Monkhorst-Pack [S10] *k*-mesh grid, based on the Perdew–Burke-Ernzerhof (PBE) [S11] exchange-correlation energy functional and the Projector Augmented-Wave [S12] pseudopotential method. The electronic band structures and projections were further computed on a finer 100×100×90 *k*-mesh grid to create interpolated Fermi surfaces near the Fermi level for evaluating band characteristics and curvatures.

The ordinary Hall coefficient of metals is connected to the geometry of the Fermi surfaces [49-51]. We plot $\gamma(s) = K(s)k_{ab}(s)$ in Eq. (10) of Ref. [49] for each piece $s$ on the Fermi surfaces in Figs. 4D-E, where $K$ is the mean curvature and $k_{ab} = \text{diag}(k_1, k_2, 0)$ is the radius-of-curvature field for the two principal Fermi-surface radii of curvature $k_1$ and $k_2$. We depict the representative electron orbits for the FS3 in Figs. S7A-B that give the electron-type Hall contribution in the effective two-band model (Eq. (5)). The contribution to the low-field Hall effect is captured by the so-called *l*-curve introduced in Ref. [51] as shown in Figs. S7C-D. Here, *l* is the mean free path vector ($\boldsymbol{l} = \boldsymbol{v}_F \tau$), and we set the relaxation time $\tau(s)$ at each Fermi surface piece $s$ as a constant. The loop for the orbit 1 (Fig. S7C) encloses the area of the *l*-space counter-clockwisely behaving as electron-type carriers. For the orbit 2 (Fig. S7D), the electron-type contribution from the three branches exceeds the hole-type area at the center of the loop resulting in the electron-type carriers in total.

**S6. Simulation curves by using different parameters.**

In the analysis in the main text, we optimize the two-band model parameters by minimizing the residual sum of squares. In order to simultaneously optimize $\sigma_{xx}(B)$ and $\sigma_{xy}(B)$, we use the weighed evaluation function, $\sum \frac{\Delta\sigma_{xx}^2}{\sigma_{xx0}^2} + \frac{\Delta\sigma_{xy}^2}{\sigma_{xy0}^2}$, where $\sigma_{xx0}$ and $\sigma_{xy0}$ are a constant with a ratio $\sigma_{xx0}/\sigma_{xy0} = 100$ since $\sigma_{xx}$ is two orders of magnitude larger than $\sigma_{xy}$. Fitting results are shown in Figs. 3A-H. We find that $n_e$ and $n_h$ are four orders of magnitude different and the concentration of the hole band is quite small (Figs. 3G-H). This is due to the simplification of the electronic structure to the two-band model, which adopt a high-mobile low density hole pocket as the second term in Eq. (5) and effectively treat large hole and electron bands regarding FS3 as a single large electron pocket for the first term in Eq. (5). To see the necessity of the high-mobile hole band for the robustness of the two-band fitting, we fit the data with a constraint so that $n_e$ and $n_h$ are a comparable order of magnitude. Even with a constraint on these parameters, we could not find a good fit when $n_e$ and $n_h$ were the same order of magnitude. To reproduce the negative slope of $\sigma_{xy}$ at higher fields (Figs. 3C, 3F), we need to set $n_e \sim 50\times10^{20}$ cm$^{-3}$, one order of magnitude larger than $n_h \sim 6\times10^{20}$ cm$^{-3}$ at 10 K. These choices give curves of $\sigma_{xx}$ and $\sigma_{xy}$ at 10 K (see Figs. S8A-B) that show good agreement, but the fit with $\sigma_{xx}$ is relatively worse than that in Fig. 3B. Furthermore, the agreement in $\sigma_{xx}$ and $\sigma_{xy}$ at 1.8 K (Figs. S8C-D) becomes worse, especially in the kagome state region, which shows a positive slope in field (see red curve Fig. S8D), opposite to that observed. We interpolate the fitting curves to the IP1-2 state regions, which deviate from the observation more significantly than shown above in Fig. S6C. Figures S8E-F are the fitting parameters at each temperature. The $n_e$ shows a large variation with temperatures (see Fig. S8E), where the value is halved from $T = 10$ K to 1.8 K. This is an unreasonable behavior as the carrier density in metals. From these results, we conclude that the failure of this fit is due to avoiding high-mobile hole pockets.

**S7. Single crystal x-ray diffraction for ErGa$_2$.**

The crystal structure was investigated by a single-crystal x-ray diffractometer at the synchrotron facility SPring-8. We performed synchrotron x-ray diffraction experiments on BL02B1 at SPring-8 in Japan [S13] using a high-quality single crystal of ErGa$_2$. A two-dimensional detector CdTe PILATUS, which had a dynamic range of $\sim 10^7$, was used to record the diffraction pattern. Diffraction data collection for crystal structural analysis was performed using a RIGAKU RAXIS IV diffractometer. Intensities of equivalent reflections were averaged and the structural parameters were refined by using Jana2006 [S14].

Figure S9 shows x-ray diffraction data of ErGa$_2$ at 300 K. The Bragg peaks are isotropic and there are no domains with different orientations. The results of the structural analysis of ErGa$_2$ at 300 K are summarized in Table S1. The obtained crystal structure is consistent with previous reports [18].

**References and Notes**

S1. J. R. Schrieffer, and P. A. Wolff, Relation between the Anderson and Kondo Hamiltonians, Phys. Rev. **149**, 491 (1966).

S2. J. R. Anderson, A. V. Gold, de Haas-van Alphen effect and internal field in iron, Phys. Rev. Lett. **10**, 227 (1963).

S3. R. C. Young, R. G. Jordan, and D. W. Jones, de Haas-van Alphen effect in gadolinium, Phys. Rev. Lett. **31**, 1473 (1973).

S4. J. Rossat-Mignod et al., Magnetic properties of cerium monopnictides, J. Magn. Magn. Mater. 31-34, 398 (1983).

S5. J. H. van Vleck, on the theory of antiferromagnetism, J. Chem. Phys. **9**, 85 (1941).

S6. H. Ishizuka, and N. Nagaosa, Spin chirality induced skew scattering and anomalous Hall effect in chiral magnets, Sci. Adv. **4**, eaap9962 (2018).

S7. Y. Fujishiro et al., Giant anomalous Hall effect from spin-chirality scattering in a chiral magnet, Nat. Commun. **12**, 317 (2021).

S8. G. Kresse, and J. Furthmüller, Efficient iterative schemes for ab initio total-energy calculations using a plane-wave basis set, Phys. Rev. B **54**, 11169 (1996).

S9. G. Kresse, and J. Furthmüller, Efficiency of ab-initio total energy calculations for metals and semiconductors using a plane-wave basis set, Comput. Mater. Sci. **6**, 15 (1996).

S10. H. J. Monkhorst, and J. D. Pack, Special points for Brillouin-zone integrations, Phys. Rev. B **13**, 5188 (1976).

S11. J. P. Perdew, K. Burke, and M. Ernzerhof, Generalized Gradient Approximation Made Simple, Phys. Rev. Lett. **77**, 3865 (1996).

S12. P. E. Blöchl, Projector augmented-wave method, Phys. Rev. B **50**, 17953 (1994).

S13. K. Sugimoto, H. Ohsumi, S. Aoyagi, E. Nishibori, C. Moriyoshi, Y. Kuroiwa, H. Sawa, and M. Takata, Extremely high resolution single crystal diffractometry for orbital resolution using high energy synchrotron radiation at Spring-8, AIP Conf. Proc. **1234**, 887 (2010).

S14. V. Petříček, M. Dušek, and L. Palatinus, Crystallographic computing system JANA2006: general features, Z. Kristallogr. Cryst. Mater. **229**, 345 (2014).

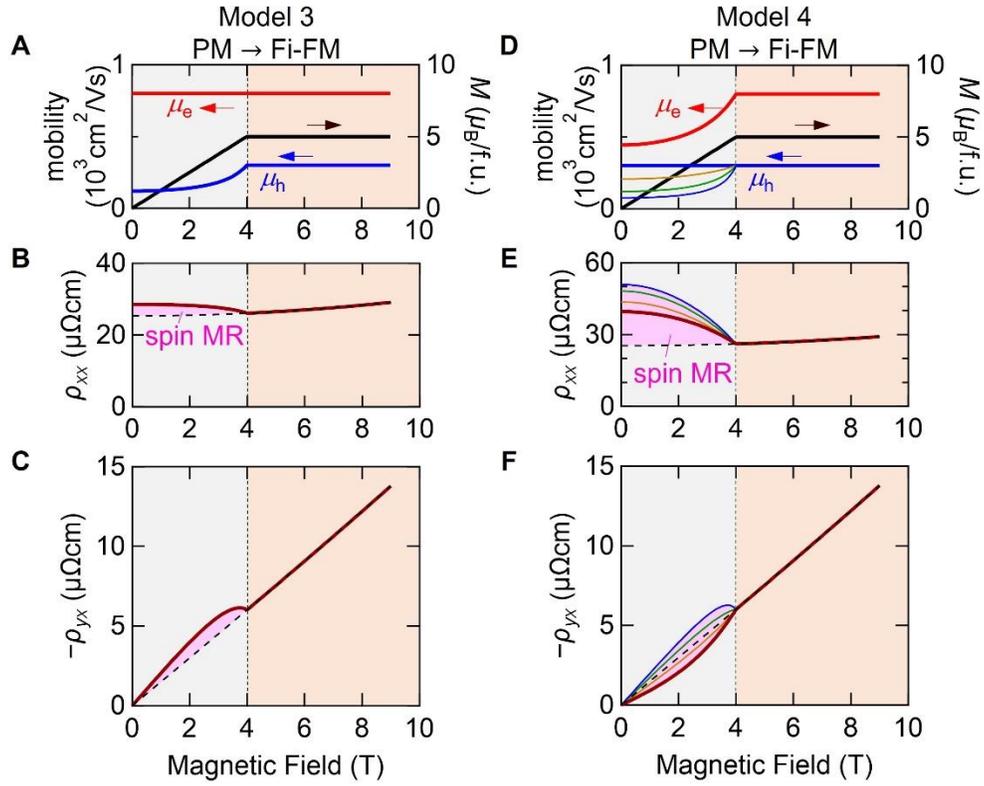

**FIG. S1. Variation of the MM Hall effect.**

**A-C** Calculated field dependence of **A** magnetization ($M$) and electron ($\mu_e$) /hole ($\mu_h$) mobilities, **B** resistivity ($\rho_{xx}$), **C** Hall resistivity ($\rho_{yx}$) for Model 3, where the mobility change exclusively in the hole carriers as $1/\tau^h_{mag} \propto 1-M^2/M_0^2$. **D-F** Corresponding plots for Model 4, where both carriers are affected by spin-disorder scattering. Thin yellow, green, blue lines are the cases when the magnitude of $1/\tau^h_{mag}$ are considered as shown by $\mu_h$ in **D**.

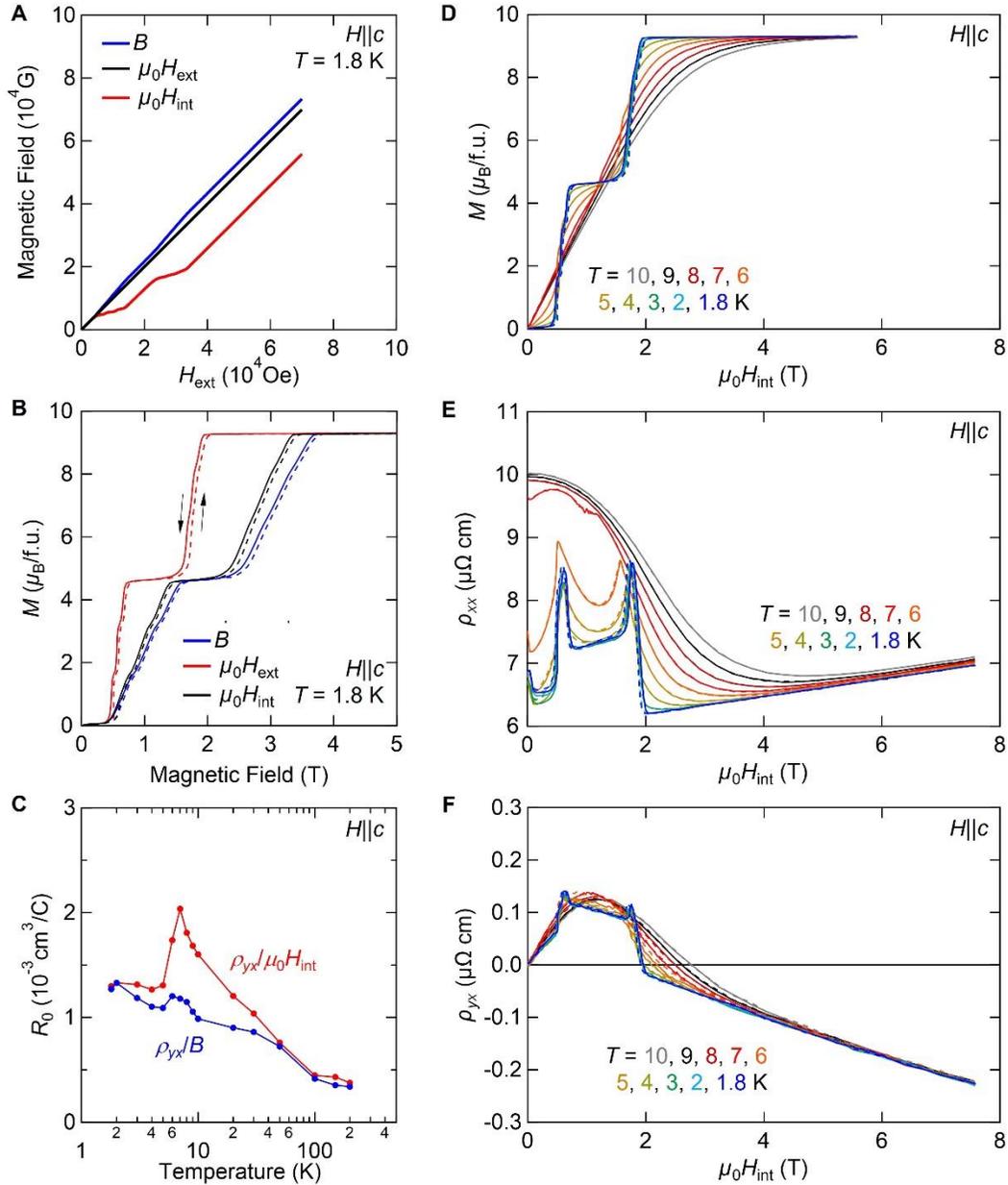

**FIG. S2. Demagnetization correction of the field-dependences of $M$, $\rho_{xx}$, and $\rho_{yx}$.**

**A** Evolution of the magnetic induction $B$ (= $\mu_0 H_{ext} - N_d M + M$), the internal magnetic field $\mu_0 H_{int}$ (= $\mu_0 H_{ext} - N_d M$), and the external field $\mu_0 H_{ext}$ at $T = 1.8$ K as the function of $\mu_0 H_{ext}$ for $H||c$. **B** Evolution of $M$ at $T = 1.8$ K for $H||c$, where various horizontal axes, $B$, $\mu_0 H_{int}$, and $\mu_0 H_{ext}$, are used. Solid (dashed) lines are for the field-decreasing (-increasing) scans.

**C** Temperature dependence of the Hall coefficient ($R_0$) for different magnetic fields. **D-F** $\mu_0 H_{int}$ dependence of $M$, $\rho_{xx}$, and $\rho_{yx}$ at various temperatures for $H||c$. Solid (dashed) lines are for the field-decreasing (-increasing) scans.

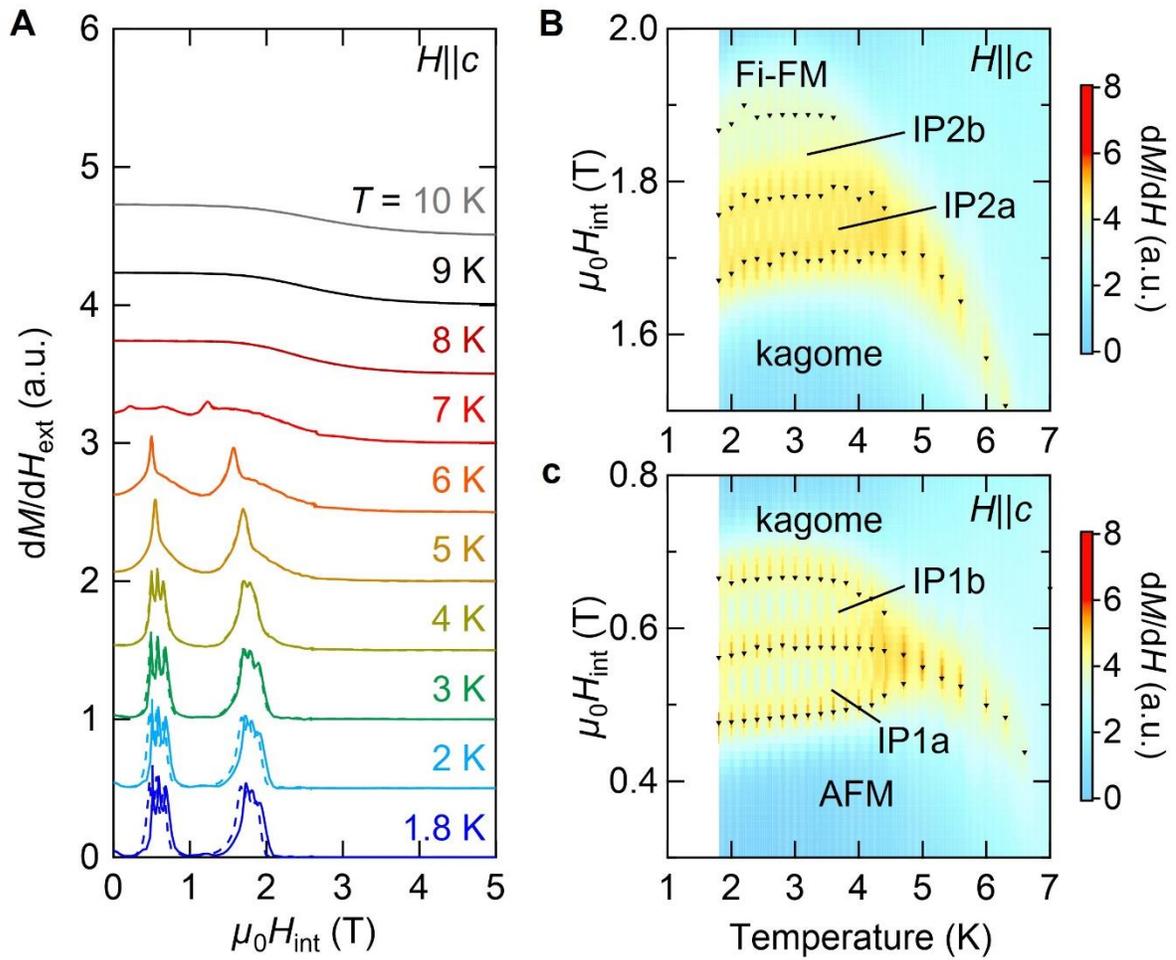

**FIG. S3. Multistep metamagnetism and phase diagrams of ErGa$_2$.**

**A** $\mu_0 H_{int}$ dependence of the field derivative of $M$ (d$M$/d$H_{ext}$) at various temperatures for $H\|c$. Solid (dashed) lines are for the field-decreasing (-increasing) scans. **B-C** the color map of d$M$/d$H$ of ErGa$_2$ for $H\|c$. The $H$-$T$ regions for the intermediate states (IP1a-b and IP2a-b, see SI text) are shown in an expended scale. Black triangles are the positions of d$M$/d$H$ peaks identified by the field-decreasing scan in Fig. S3A.

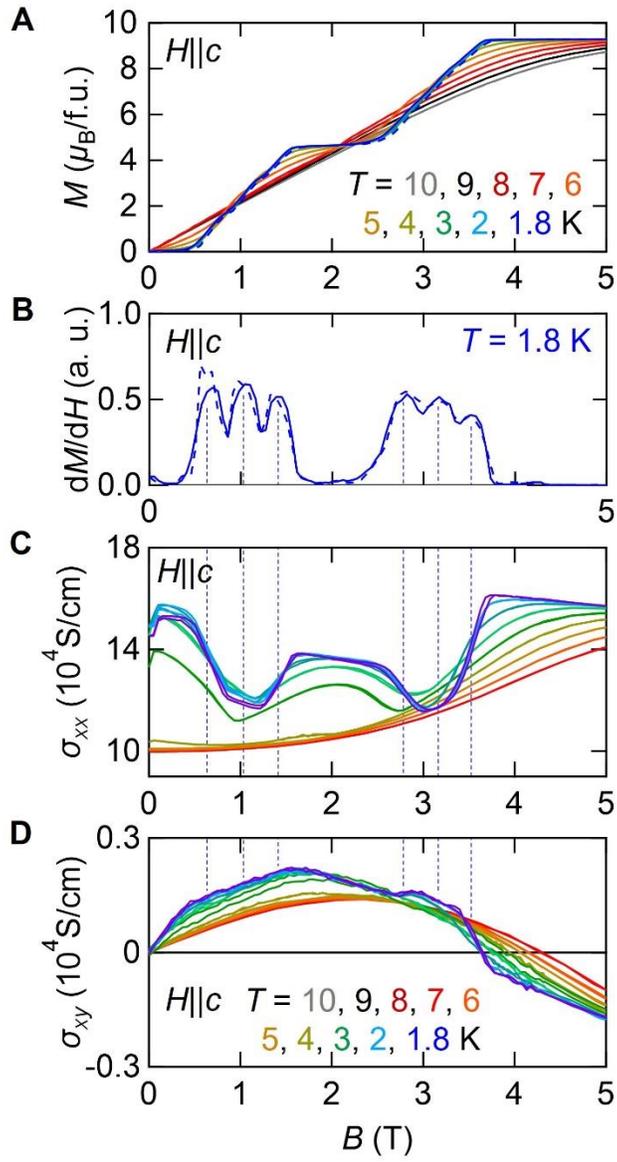

**FIG. S4. Field-dependence of *M* and magnetotransport tensors.**

**A-D** *B*-dependences of *M*, d*M*/d*H*, $\sigma_{xx}$, and $\sigma_{xy}$ at various temperatures for *H*||*c*. Vertical dashed lines are guides to the eye for the d*M*/d*H* peak positions.

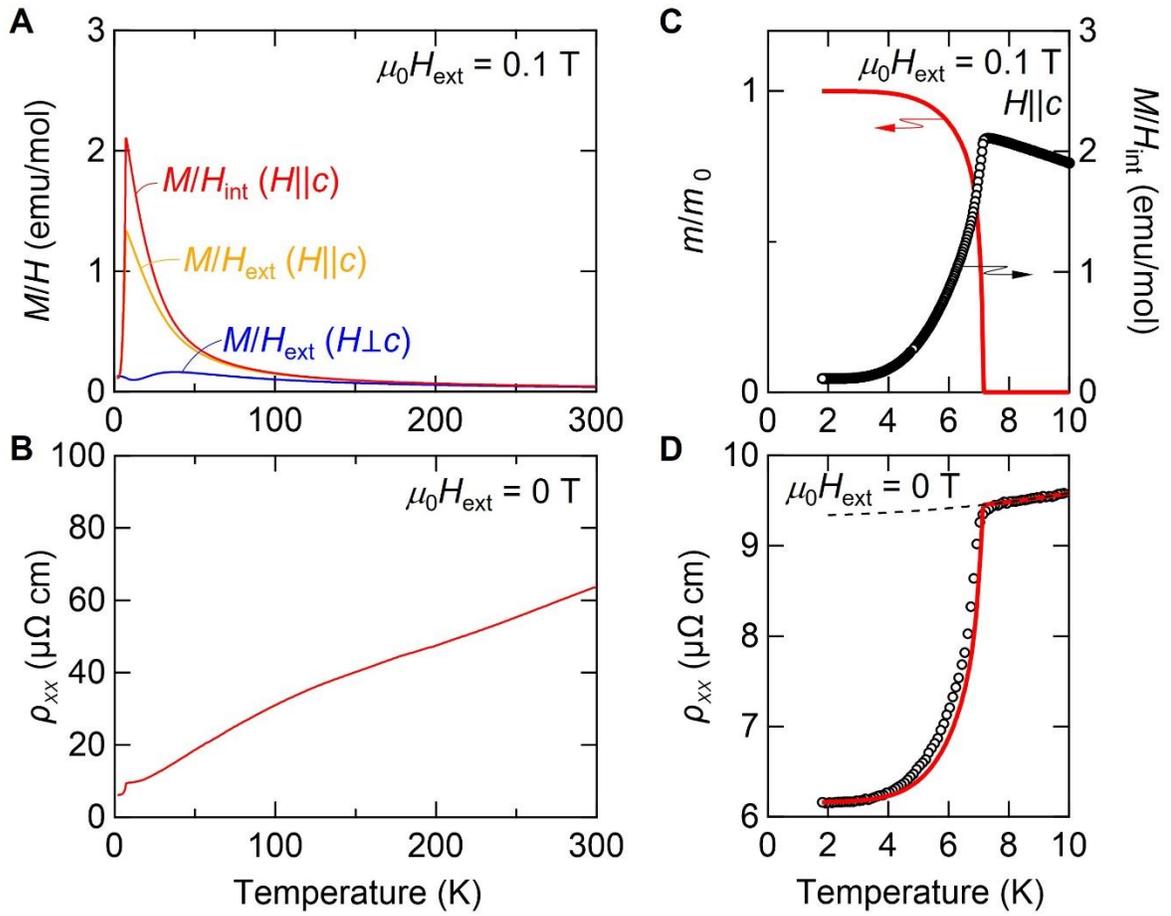

**FIG. S5. Spin-disorder scattering in AFM state.**

**A** Temperature dependence of magnetic susceptibility ($M/H$) of ErGa$_2$ for $H\|c$ (orange) and $H\perp c$ (blue) at $\mu_0 H_{ext} = 0.1$ T. For $H\|c$, this is demagnetization effect corrected to obtain $\chi_{\|c}$ (= $M/H_{int}$) in Eq. (S1) as shown by red curve. **B** Temperature dependence of resistivity for $J\perp c$ at zero field. **C** Normalized order parameter $m/m_0$ (red line, left axis) obtained by Eq. (S2) with $\chi_{\|c} = M/H_{int}$ (open black circle, right axis) **D** The simulated $\rho_{xx}$ (red) and the observed $\rho_{xx}$ (open circle). The dashed curve represents $\rho_0 + \rho_{ph}$ (see text).

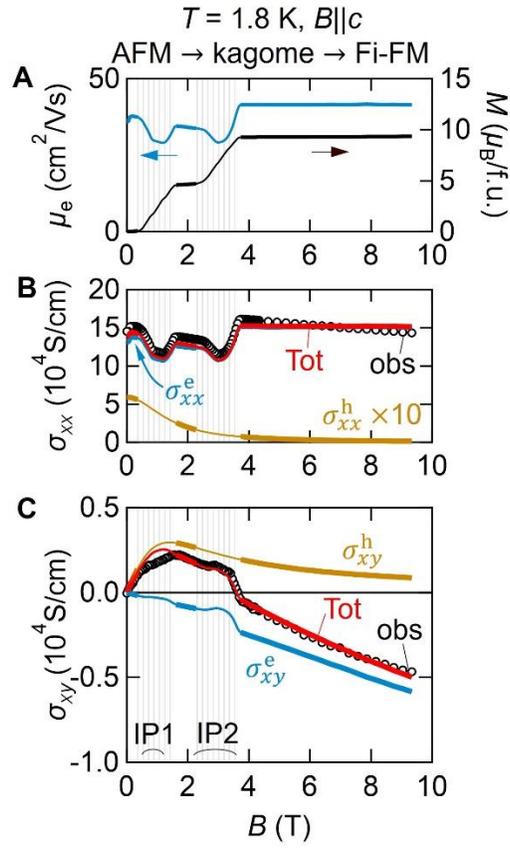

**FIG. S6. Interpolation of field-dependent $\mu_e$ into the IP state regions.**

**A-C** Corresponding figures to Figs. 3D-F by interpolating the assumption, $\rho_{xx}(B)/\rho_{xx0} \propto 1/\mu_e(B)$, to the IP state regions.

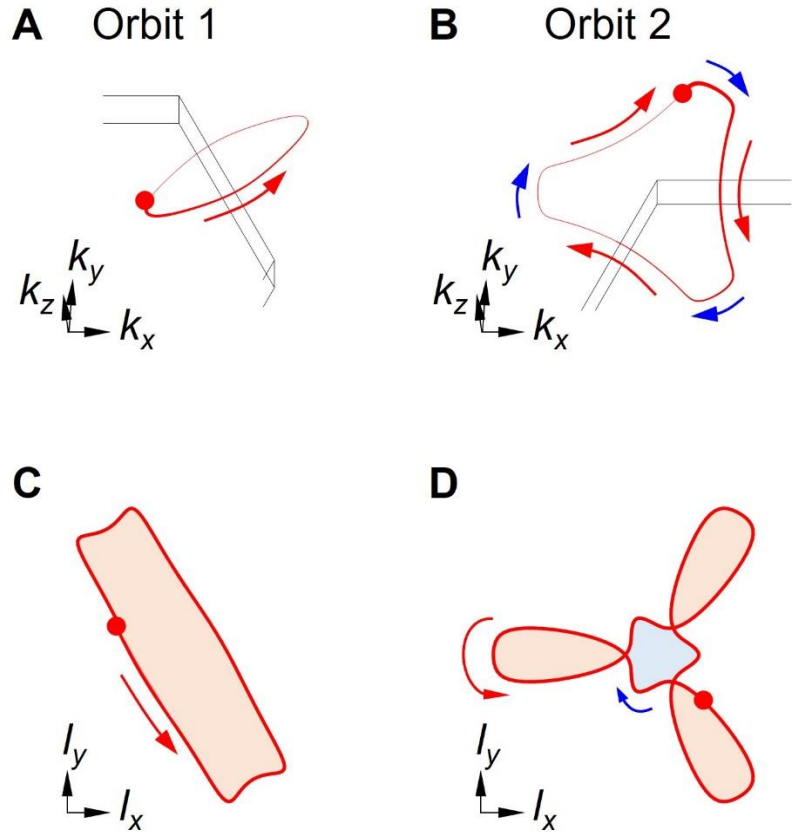

**FIG. S7. Electron-type orbits in the low-field Hall effect.**

**A (B)** The representative reciprocal-space electron orbit 1 (2) contributing to the $n_e$ of the two-band model in Eq. (5). Arrows are the direction of motion under $B\|c$ based on the semiclassical theory of electron dynamics. Red (blue) arrow represents the counter-clockwise (clockwise) deflection. **C-D** Corresponding *l*-curves as introduced in Ref. [51]. Red (blue) area counter-clockwisely (clockwisely) encloses the *l*-space. The red dot corresponds to the start point of the orbit in A-B.

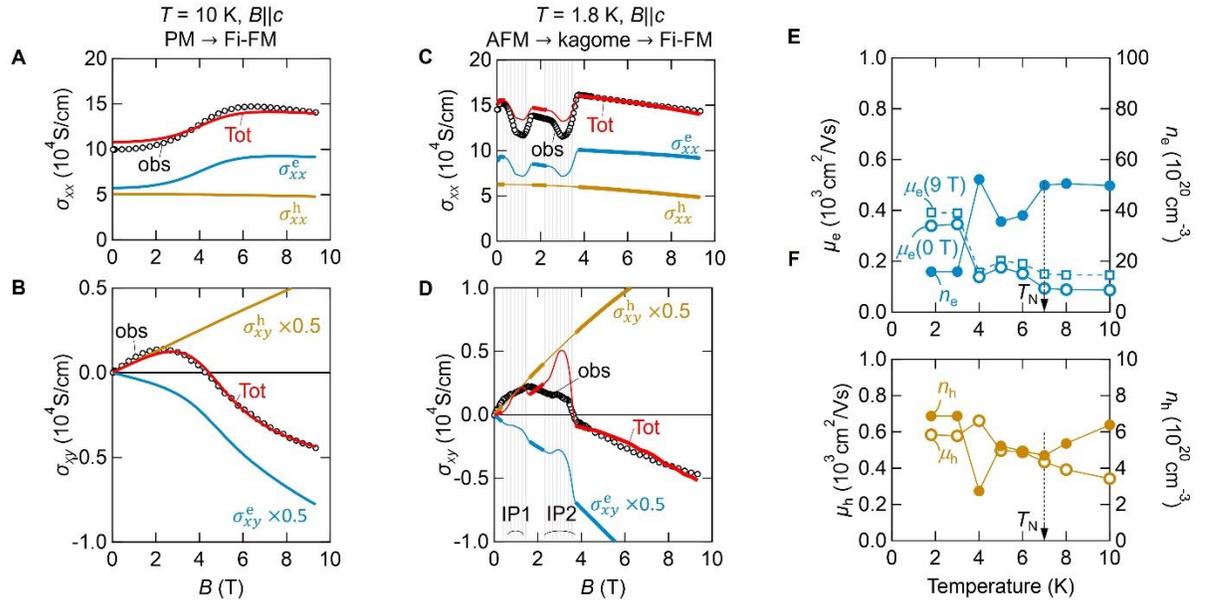

**FIG. S8. Fitting results with a constraint that $n_e$ and $n_h$ are close order of magnitudes.**

**A-F** Corresponding figures to Figs. 3B-C, E-F, G-H, respectively. A constraint is imposed so that the orders of $n_e$ and $n_h$ are closer with each other.

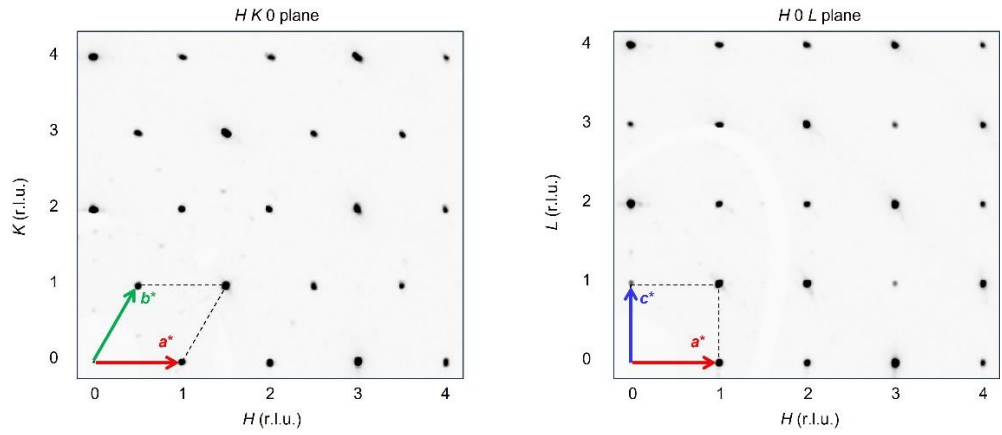

**FIG. S9. Single crystal x-ray diffraction pattern of ErGa$_2$.**

(Left) Fundamental Bragg peaks in the $H\,K\,0$ plane, and (Right) those in the $H\,0\,L$ plane.

Table S1| Structural parameters and crystallographic data of ErGa$_2$ at 300 K. The space group is $P6/mmm$ (No. 191), and $a = b = 4.1877(4)$ Å, $c = 4.0175(2)$ Å, $\alpha = \beta = 90°$, $\gamma = 120°$.

| Atom | Site | Sym. | $x$ | $y$ | $z$ | $U_{11}$ (Å$^2$) | $U_{33}$ (Å$^2$) | $U_{12}$ (Å$^2$) |
|---|---|---|---|---|---|---|---|---|
| Er | 1$a$ | 6/$mmm$ | 0 | 0 | 0 | 0.008063(1) | 0.006641(17) | 0.004031(6) |
| Ga | 2$d$ | -6/$m$2 | 2/3 | 1/3 | 1/2 | 0.006169(14) | 0.02323(6) | 0.003085(7) |

| | |
|---|---|
| Temperature (K) | 300 |
| Wavelength (Å) | 0.309561 |
| Crystal dimension ($\mu$m$^3$) | 132 x 96 x 41 |
| Space group | $P6/mmm$ |
| $a$ (Å) | 4.1877(4) |
| $c$ (Å) | 4.0175(2) |
| Z | 1 |
| $F$(000) | 130 |
| (sin$\theta$/$\lambda$)$_{max}$ (Å$^{-1}$) | 1.79 |
| $N_{Total}$ | 6434 |
| $N_{Unique}$ | 590 |
| Average redundancy | 10.905 |
| Completeness (%) | 92.19 |
| $N_{parameters}$ | 6 |
| $R_1$ ($I>3\sigma$) [number of reflections] | 1.05% [572] |
| $R_1$ (all) [number of reflections] | 1.08% [590] |
| $wR_2$ (all) [number of reflections] | 1.37% [590] |
| GOF (all) [number of reflections] | 0.99 [590] |